# Combined SANS and SAXS study of the action of ultrasound on the structure of amorphous zirconia gels

N.N. Gubanova[1], A.Ye. Baranchikov[2]*, G.P. Kopitsa[1], L. Almásy[3], B.S. Angelov[4], A.D. Yapryntsev[2], L. Rosta[3], V.K. Ivanov[2,5]

[1] *Petersburg Nuclear Physics Institute, Gatchina, Russia*

[2] *Kurnakov Institute of General and Inorganic Chemistry of the Russian Academy of Sciences, Moscow, Russia*

[3] *Institute for Solid State Physics and Optics, Wigner Research Centre for Physics, Hungarian Academy of Sciences, Hungary*

[4] *Institute of Macromolecular Chemistry, Prague, Czech Republic*

[5] *National Research Tomsk State University, Tomsk, Russia*

*Corresponding author • E-mail: a.baranchikov@yandex.ru • Postal address: Leninsky ave., 31, Moscow 119991 Russia • Tel.: +74956338534

**Highlights**

- Ultrasonication results in a dramatic increase in the surface fractal dimension of $ZrO_2 \times xH_2O$
- SANS and SAXS prove a similar effect of sonication on the mesostructure of $ZrO_2 \times xH_2O$
- Ultrasonication increases the specific surface area of $ZrO_2 \times xH_2O$

**Abstract**

In the present work, we have studied for the first time the combined effect of both sonication and precipitation pH on the structure of amorphous zirconia gels synthesized from zirconium(IV) propoxide. The techniques of small-angle neutron and X-ray scattering (SANS and SAXS) and low temperature nitrogen adsorption provided the integral data on the changes in the microstructure and mesostructure of these materials caused by ultrasonic treatment. Amorphous $ZrO_2 \cdot xH_2O$ synthesized under ultrasonic treatment was found to possess a very structured surface, characterized by the surface fractal dimension 2.9–3.0, compared to 2.3–2.5 for the non US-assisted synthesis, and it was also found to possess a higher specific surface area, while the sizes of the primary particles remain unchanged.

## 1. Introduction

The sonochemical approach is widely used for the synthesis of a great variety of advanced inorganic materials, including metal oxides and hydroxides from solutions and suspensions [1–4], in hydrothermal media [5] and in solid phase [6, 7]. The advantages of the ultrasound-assisted sol–gel technique over conventional routes of nanomaterials synthesis include shortening the synthesis duration due to faster hydrolysis, leading to more uniform particle size distribution, higher surface area, better thermal stability, and improved phase purity [4]. Examples of successful ultrasound-assisted sol–gel synthesis of metal oxide nanostructures include $TiO_2$ [8–10], ZnO [11–13], $MoO_3$ [14], $In_2O_3$ [15], $ZrO_2$ [16], $SiO_2$ [17–19], *etc*. It was shown that in a number of cases, sonochemically prepared materials demonstrate better characteristics than those synthesized by conventional methods. For example, nickel hydroxides obtained by ultrasonic-assisted techniques [20–23] possessed higher electrochemical performance. Similarly, layered double hydroxides obtained by an ultrasound-enhanced technique showed larger adsorption capacity for humic substances [24]. Recently, formation of high specific surface area porous adsorbents with the use of ultrasound was also reported [25, 26].

When power ultrasound is introduced in the liquid-based media, absorbed acoustic energy gives rise to a number of physical effects, resulting in turbulent fluid movement in the vicinity of cavitational bubbles [27–29]. Near the phase boundary (e.g. solid-liquid interface), high speed microjets and shockwaves are formed [30]. Thus, acoustic treatment of suspensions consisting of relatively large particles, of which the size is comparable to the size of a collapsing cavitation bubble (d > 0.5–1 μm), can lead to several specific effects, including de-agglomeration, decrease of mean particle size, increase of the surface area, amorphization *etc*. [31–33]. Interestingly, cavitation bubble nucleation and collapse near the solid-liquid boundary is strongly dependent on a number of factors, including the wettablility of the surface [34].

Zirconia and zirconia-based materials are of great importance because of the wide variety of their industrial applications (catalysts, oxygen-conducting materials *etc*.) [35–37]. The most convenient approach to synthesize these materials is based on the precipitation of amorphous hydrous zirconia gels in aqueous media using zirconium-containing precursors (e.g., zirconyl nitrate, zirconium alkoxides) and subsequent thermal or hydrothermal treatment of the resulting

$ZrO_2·xH_2O$ gel [38,39]. The morphology and phase composition of such synthesized zirconia is, to a large extent, governed by the structure of the precursor gel, which in turn depends on the conditions of precipitation (*e.g.* composition, temperature, acidity of starting solution, *etc.*) [39–41]. For example, variation of precipitation pH changes the relative rates of the hydrolysis and condensation of zirconium-based clusters. The excess of alkali in the reaction media results in rapid hydrolysis and condensation, forming a branched metal oxy-hydroxide network. In particular, hydrous zirconia precipitated above the point of zero charge possesses a higher specific surface area and surface fractal dimension than when it is synthesized at low pH [39]. Similar behavior was shown for gels synthesized from zirconium isopropylate [41]. High surface fractal dimensions of amorphous hydrous zirconia obtained by the sol-gel route may, in some cases, remain unchanged, even after crystallization [42].

Recently, the application of ultrasound to the synthesis of zirconia-based materials has attracted certain interest. For example, it was established that ultrasonic cavitation disaggregates the agglomerates of zirconia colloidal particles, reduces the amount of physically and chemically bound water (as well as the amount of adsorbed ions), and leads to a notable increase in the specific surface area of zirconia amorphous samples [16, 43]. Ultrasonically treated zirconia was shown to transform faster from the monoclinic to the tetragonal phase [44]. Obviously, these changes in the properties of zirconia are closely related to the effects of ultrasonication on the structure of amorphous hydrous zirconia formed during sonochemical-assisted precipitation, but these important structural aspects still remain virtually unstudied. Moreover, the corresponding experimental reports are contradictory. For instance, it was shown [45] that sonication increases the rate of linear polymeric clusters formation and notably decreases the gyration radius of $ZrO_2$ particles. Contrariwise, our later experiments [46, 47] revealed the increase in the surface fractal dimension and in the size of individual particles in sonochemically prepared amorphous zirconia gels [47]. Up to now, studies on the fractal structure of gels forming under the action of ultrasound were conducted primarily for hydrous silica [48, 49]. For example, Vollet *et al.* [49] have shown that ultrasound-stimulated wet silica gels possessed a lower fractal dimension than conventional ones. However, upon drying the former has shown a higher pore volume and specific surface area.

In the present work, we have studied for the first time the combined effect of both sonication and precipitation pH on the structure of amorphous zirconia gels precipitated from zirconium(IV) propoxide. Our investigation was primarily based on the use of small-angle neutron scattering (SANS) and X-ray scattering (SAXS), these being the most suitable methods for the structural study of amorphous materials. SANS, SAXS and low temperature nitrogen adsorption provided the integral data on the changes in the microstructure and mesostructure of these materials caused by ultrasonic treatment.

## 2. Experimental
### 2.1 Synthesis of samples

Hydrous zirconia xerogels were prepared as follows. First, nitric acid or aqueous ammonia was added to distilled water, giving 55 ml portions of solutions adjusted to pH 2.66, 5.46, 6.25, 8.26, 11.30. Then, 4 ml of 70 wt.% zirconium propoxide solution in propanol (Aldrich, 333972) was added dropwise to each portion under constant stirring. The resulting precipitates have been stirred for 30 minutes, then thoroughly washed with distilled water and dried in air at 150°C overnight.

Ultrasound-assisted synthesis was conducted according to a similar procedure. Starting solutions were adjusted to pH 5.46, 6.25, 8.26, 11.30, and sonicated using a Bandelin Sonopulse 3200 generator (titanium horn SH 213 G with TT13 tip) throughout the precipitation (15 min) and subsequent stirring (30 min) of resultant suspensions. The output ultrasonic specific power measured using a standard calorimetric technique [50, 51] was equal to 13±1 W/cm$^2$. The TT13 tip was immersed 10 mm below the surface of the solutions. To prevent overheating the whole process was conducted in a thermostated cell at 25°C.

For the sake of clarity, hydrous zirconia xerogels synthesized from pH 2.66, 5.46, 6.25, 8.26, 11.30 solutions are hereafter named Z-2X, Z-5X, Z-6X, Z-8X, Z-11X, where X indicates whether the synthesis was ultrasonically assisted (X=U) or not (X=C).

### 2.2 Methods of analysis
#### 2.2.1 Thermal analysis, XRD, low-temperature nitrogen adsorption and SEM study of the samples

Thermal analysis (TGA/DTA) was performed using a NETZSCH STA 409 PC/PG instrument in the temperature range 20-900°C in air (heating rate $\beta = 10$°/min, platinum crucibles, ~30 mg samples). Mass-spectral data during thermal analysis were collected by QMS 403 C Aëolos®. X-ray diffraction (XRD) patterns were recorded using a Rigaku D/MAX 2500 diffractometer (CuK$_\alpha$ radiation) over a 2θ range of 10–85 ° with an increment 0.02 °/step at the rate of 2 °/min. Low temperature nitrogen adsorption measurements were conducted using an ATX-6 analyzer (Katakon, Russia). Before the measurements the samples were outgassed at 150 °C for 30 min under a dry helium flow. Determination of the surface area was carried out by the 8-point Brunauer-Emmett-Teller (BET) method. The microstructure of powders was also studied using a Carl Zeiss NVision 40 scanning electron microscope with a Schottky field emission Gemini column, operated at 1 kV acceleration voltage.

### 2.2.2 Neutron and X-ray scattering measurements

The SANS experiment was performed at the "Yellow submarine" instrument of the BNC research reactor in Budapest (Hungary). The use of two neutrons wavelengths ($\lambda = 0.46$ and 1.2 nm), and two sample-to-detector distances (1.33 and 5.6 m) provided measurements in a wide momentum transfer range ($4.2 \cdot 10^{-2} < q < 3.8$ nm$^{-1}$, where: $q = 4\pi\lambda^{-1}\sin(\theta/2)$, $\theta$ is the scattering angle). The scattered neutrons were detected by a two-dimensional position-sensitive BF$_3$ gas detector (64×64 cells of 1 cm × 1 cm).

Samples of amorphous zirconia xerogels were placed in 1 mm thick quartz cells. Apparent density $\rho_H$ of each sample was calculated as the mass of powder divided by its volume. The raw data were corrected using the standard procedures [52], taking into account the scattering from the set-up equipment and cell. The resulting 2D isotropic spectra were averaged azimuthally and their absolute values were determined by normalizing them to the incoherent scattering cross-section of water. All the measurements were done at room temperature. The BerSANS software [53] was used for data processing.

A small-angle X-ray scattering (SAXS) experiment was performed using a pinhole camera (Molecular Metrology SAXS System) attached to a microfocused X-ray beam generator (Bede, Durham, UK) operating at 45 kV and 0.66 mA (30 W). The camera was equipped with a multiwire, gas-filled area detector with an active area diameter of 20 cm and 512×512 pixels (Gabriel design). An X-ray diode was put as a beamstop in the center of the detector. Measurements were conducted in a momentum transfer range $5 \cdot 10^{-2} < q < 11$ nm$^{-1}$ ($\lambda = 0.154$ nm). The scattering intensities were put on an absolute scale using a glassy carbon standard.

## 3. Results and discussion

X-ray powder diffraction patterns of all the gels are typical to hydrous zirconia precipitated from aqueous media at ambient temperature, and show two extremely broad peaks at ~30 and ~50–60°2θ. The nature of these peaks was extensively discussed elsewhere [39, 54]. Thus, all the samples obtained in this work are of an amorphous nature.

Thermograms of all the samples are almost identical, giving a total weight loss of 77–82%, which is related to the physically and chemically bonded water. Taking into account thermal analysis data, and supposing that the entire weight loss is connected with the elimination of water, we conclude that the chemical composition of xerogels meets the approximate formula $ZrO_2 \times 1.7H_2O$. Crystallization of zirconia occurs at ~430°C and is accompanied by a pronounced exothermic effect. Note that crystallization of $ZrO_2$ from amorphous xerogels precipitated under acidic conditions is often accompanied by decomposition of various impurities (*e.g.*, nitrates), which gives a marked weight loss at crystallization temperature [39, 55]. The absence of this effect allows the conclusion that xerogels do not contain notable amounts of adsorbed species. To confirm this, we used thermal analysis combined with mass-spectroscopy of gases, evolving during samples decomposition. At the low temperature stage of weight loss (~100°C) for Z-2C and Z-5C samples, we have observed only weak signals corresponding to the elimination of nitrous gases ($NO_2$, $NO$ and $N_2O$), which points out that only a small quantity of nitric acid is adsorbed on the surface of these xerogels, which is easily eliminated.

Specific surface area measurements data are presented in Table 1. One can see that for a control series of samples, the increase in pH of the starting solutions does not result in a significant change in specific surface area or specific pore volume. It has been previously shown that precipitation of hydrous zirconia from inorganic salts in alkaline solutions (pH > 6) gives powders with a developed surface ($S_{BET}$ > 100 $m^2 \times g^{-1}$), while precipitation in acidic media results in dense powders with a sufficiently lower surface area [39, 56]. Note that the synthetic route used here leads to the formation of powders with a developed surface, regardless of the acidity of the starting solution. The reason for such a result is probably related to the high degree of supersaturation in the system when zirconium alkoxide is hydrolyzed in aqueous media.

High surface area values are also typical for all the samples synthesized under sonication; however, in this case an increase in precipitation pH results in a slight decrease in the specific surface area of precipitated $ZrO_2 \times xH_2O$.

Data presented in Table 1 indicate that the application of ultrasound results in the formation of xerogels with a slightly higher surface area and pore volume. This is probably caused by the action of microjets of liquid which arise when cavitation bubbles collapse near the interphase boundaries (*i.e.,* the surface of solid particles is suspended in aqueous media).

**Table 1.** Specific surface area and specific pore volume estimated from nitrogen adsorption data

| Sample name | Specific surface area, | Specific pore volume, | Sample name | Specific surface area, | Specific pore volume, |

|       | m²×g⁻¹  | cm³×g⁻¹     |       | m²×g⁻¹ | cm³×g⁻¹     |
|-------|---------|-------------|-------|--------|-------------|
| Z-5C  | 210±10  | 0.048±0.06  | Z-5U  | 250±10 | 0.064±0.05  |
| Z-6C  | 220±10  | 0.051±0.05  | Z-6U  | 250±10 | 0.056±0.05  |
| Z-8C  | 200±10  | 0.046±0.07  | Z-8U  | 230±10 | 0.053±0.06  |
| Z-11C | 210±10  | 0.048±0.05  | Z-11U | 220±10 | 0.052±0.06  |

Diffraction and adsorption-based methods cannot provide comprehensive information on the structure of amorphous materials. The size of the individual particles and their aggregates, as well as the structure of their surface, can be obtained using SANS and SAXS methods, which are widely used to access the mesostructure of various materials in the 1–100 nm scale range.

Figure 1 shows the experimental curves of the differential neutron cross-section $d\Sigma(q)/d\Omega$ versus momentum transfer $q$ for amorphous hydrous zirconia gels precipitated at different pH values under sonication (Fig. 1A), and without ultrasonic treatment (Fig. 1B) (For clarity of the image, the curves are shifted vertically). According to this figure, the scattering curves have the same overall shape for all samples, and display three characteristic $q$-ranges. Figure 1 also shows that scattering patterns of zirconia gels synthesized by hydrolysis of zirconium $n$-propoxide are generally similar to spectra obtained for amorphous $ZrO_2 \cdot xH_2O$ precipitated from aqueous solutions of zirconium nitrate [39].

In the range $q < 0.8$ nm⁻¹, the scattering cross-section $d\Sigma(q)/d\Omega$ for all the samples satisfies the power law $q^{-n}$. Such power-law dependence is observed for a wide size distribution of scattering inhomogeneities between characteristic sizes $R_{min}$ and $R_{max}$, satisfying the condition:

$$R_{max}^{-1} << q << R_{min}^{-1} \qquad (1)$$

In addition, the power-scattering law means that inhomogeneities making the dominant contribution to scattering are sufficiently large, so that $q_{min} \cdot R >> 1$. Bale and Schmidt [57] proposed a more exact criterion for a certain characteristic size of inhomogeneities: $q_{min} \cdot R \approx 3.5$. In our case, $q_{min} = 4.2 \cdot 10^{-2}$ nm⁻¹ and the characteristic size of the inhomogeneities, which can be detected on the instrument used, is equal to $R \approx 80$ nm.

The absence of the Guinier region on the scattering curves for low $q$ values ($< 4.2 \cdot 10^{-2}$ nm⁻¹) means that the upper self-similarity limit $\xi$ of the surface fractal is larger than the maximum size $R_{max}$ of the inhomogeneities that can be detected in the experiment with a given resolution, *i.e.,* higher than 80 nm.

Bale and Schmidt [57] have also shown that for porous materials in which the interfaces are surface fractals with a fractal dimension $2 \leq D_s < 3$, the scattering intensity depends on the momentum transfer in a power law form:

$$I(q) = A \cdot q^{-n} \qquad (2)$$

in the $q > 1/\xi$ momentum-transfer region, where $3 < n = 6 - D_s \leq 4$, and $\xi$ is the upper self-similarity limit of the length range in which the internal surface is fractal. Thus, the fractal dimension $D_S$ can be obtained from the slope of the straight-line sections of the experimental curves $I(q)$ plotted in log–log scale. Parameter $A$ is a power-law prefactor, which is related to the interface surface $D_s$ [57].

When the upper self-similarity limit $\xi$ of the fractal is not larger than the maximum size $R_{max}$ of the inhomogeneities (pores) that can be detected in the scattering experiment with a given resolution, the estimation of $\xi$ can be obtained from the analysis of the scattering in the Guinier region $(1 < q\xi)$ [58]:

$$I(q) = G \cdot \exp\left(-\frac{q^2 R_g^2}{3}\right) \qquad (3)$$

where $G$ is the Guinier prefactor, which is in direct proportion to the product of the number of the inhomogeneities (pores) in scattering volume and the density of the neutron-scattering amplitude $\rho$ on them; $R_g$ is the radius of gyration for such inhomogeneities (pores) that can be calculated from the curve's slope of $\ln(I(q))$ versus $q^2$ and is related for spheres to the geometrical radius by $R_g = \sqrt{3/5} R_c$ [59].

The exponent $n$ values found from the slope of the straight-line sections of the experimental curves $d\Sigma(q)/d\Omega$ plotted in log–log scale lie in the range from 3.44 to 3.65 for the control samples (Fig. 1B), and from 3.04 to 3.14 for the sonicated samples (Fig. 1A), respectively. As mentioned above, if the exponent value is in the range $3 < n \leq 4$, scattering from the samples occurs on the fractal surface with the dimension $2 \leq D_{s1} = 6 - n < 3$. The absence of the Guinier region on the scattering curves for low $q$ values means that the upper self-similarity limit $\xi$ of the surface fractal is larger than the maximum size $R_{max}$ of the inhomogeneities that can be detected in the experiment with a given resolution.

In the region 0.8 < $q$ < 2 nm$^{-1}$ the so-called "shoulder of the curve" is observed, indicating the presence of small inhomogeneities (primary particles) of characteristic size $r_c$. In this case, the observed scattering depends on the shape and size $r_c$ of these particles and exhibits Guinier-like behavior. The information on the structure of their surface can be derived from the analysis of the scattering at the large $q$ region ($q$ > 2 nm$^{-1}$), where the behavior of the cross-section $d\Sigma(q)/d\Omega$ also satisfies the power law $q^{-m}$. The exponent $m$ values found from the slope of the straight-line parts of the curves plotted in log-log scale lie in the range from 3.3 to 3.86 for the control samples (Fig. 1B), and from 3.24 to 3.9 for the sonicated samples (Fig. 1A), respectively. This also corresponds to the scattering from the fractal surface with the dimension $D_{s2}$.

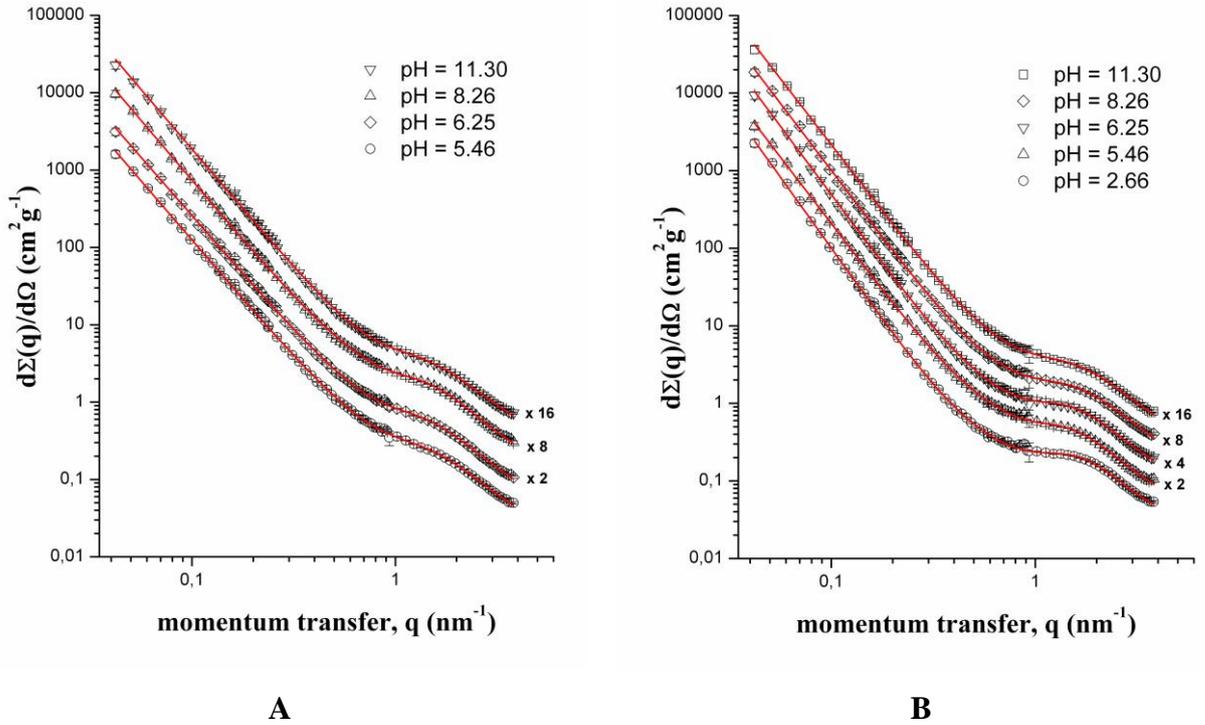

A                                      B

Fig. 1. SANS differential cross-section $d\Sigma(q)/d\Omega$ for the samples of amorphous zirconia precipitated under the action of ultrasound (A) and without it (B). Fits of experimental data by equation (8) are shown as solid lines. For the sake of clarity, cross-section values for some samples were multiplied by 2, 4, 8 and 16 (corresponding factors are given next to the curves).

Thus, the observed patterns are typical for scattering from porous systems with fractal phase boundaries [57]. They indicate also that zirconia xerogels contain two types of scattering inhomogeneities with strongly different characteristic scales. It can be concluded that these xerogels are composed of aggregates with a strongly developed fractal surface that are formed from the primary particles which, in turn, also have a fractal surface. The upper self-similarity limit $\xi$ of the large-scale aggregates is larger than ~80 nm and the lower self-similarity limit is determined by the characteristic size $r_c$ of the primary particles (*ca.* 2.5 nm).

In view of this circumstance, we used the following expression to analyze the scattering from all samples over the entire $q$ range [60]:

$$\frac{d\Sigma(q)}{d\Omega} = \frac{A_1(D_{S1})}{q^n} + G_p \cdot \exp(-\frac{q^2 r_g^2}{3}) + \frac{A_2(D_{S2})}{\hat{q}^m} + I_{inc} \quad (4)$$

where amplitudes $A_1(D_{s1})$ and $A_2(D_{s2})$ are power-law prefactors, which depend on the fractal dimensions of the aggregates and primary particles, respectively. Amplitude $G_p$ is the Guinier prefactor for the scattering from the primary particles; $\hat{q} = q/[erf(qr_g/6^{1/2})]^3$ is the momentum $q$ transferred, normalized to an error function $erf(x)$. Such a procedure allows the scattering cross-section $d\Sigma(q)/d\Omega$ to be described correctly in the intermediate region between $qr_c < 1$ (Guinier approximation) and $qr_c \gg 1$ ($q^{-m}$ asymptotics), where scattering from both surface and monodisperse inhomogeneities of characteristic size $r_c$ contributes. The constant $I_{inc}$ is independent of $q$ and is associated with the incoherent scattering on hydrogen atoms in the xerogels.

The expression (Eq. (4)) was convolved with the instrumental resolution function which is approximated by a Gaussian [61]. The experimental curves of the differential cross-section $d\Sigma(q)/d\Omega$ versus $q$ were processed by the least mean squares method over the entire measured $q$ range. The results of this analysis are given in Figs. 1 and 2, as well as in Table 2.

**Table 2.** Structural parameters of sonicated and non-sonicated zirconia gels, as obtained from SANS and SAXS measurements.

| Parameter | Sample designation | | | | | | | | |
|---|---|---|---|---|---|---|---|---|---|
| | Z-2C | Z-5C | Z-5U | Z-6C | Z-6U | Z-8C | Z-8U | Z-11C | Z-11U |
| SANS | | | | | | | | | |
| $A_1 \times 10^{-2}$, cm$^2$g$^{-1}$ | 2.4±0.3 | 3.0±0.4 | 10.9±1.1 | 3.7±0.2 | 12.4±1.1 | 4.5±0.3 | 8.0±0.6 | 5.2±0.5 | 8.9±0.8 |
| $D_{S1}$ | 2.35±0.04 | 2.43±0.03 | 2.92±0.04 | 2.46±0.03 | 2.96±0.04 | 2.53±0.04 | 2.92±0.04 | 2.56±0.04 | 2.86±0.04 |
| $G$, cm$^2$g$^{-1}$ | 0.48±0.01 | 0.65±0.01 | 0.88±0.01 | 0.60±0.01 | 1.08±0.01 | 0.56±0.01 | 0.76±0.01 | 0.58±0.01 | 0.74±0.01 |
| $r_C$, Å | 23.7±0.3 | 25.4±0.4 | 24.3±0.5 | 24.7±0.3 | 26.5±0.6 | 24.4±0.4 | 25.1±0.4 | 23.6±0.4 | 24.5±0.5 |
| $A_2$, cm$^2$g$^{-1}$ | 2.6±0.4 | 2.0±0.3 | 1.7±0.3 | 2.3±0.3 | 1.6±0.2 | 1.6±0.2 | 2.0±0.3 | 1.7±0.3 | 1.6±0.3 |
| $D_{S2}$ | 2.14±0.06 | 2.32±0.06 | 2.58±0.07 | 2.21±0.06 | 2.76±0.06 | 2.70±0.07 | 2.10±0.06 | 2.61±0.07 | 2.43±0.07 |
| $I_{inc} \times 10^{-2}$, cm$^2$g$^{-1}$ | 3.6±0.2 | 3.3±0.2 | 3.0±0.2 | 3.4±0.2 | 3.2±0.3 | 2.8±0.3 | 2.8±0.1 | 2.8±0.3 | 3.0±0.2 |
| SAXS | | | | | | | | | |
| $D_{S1}$ | 2.17±0.02 | 2.34±0.02 | 2.53±0.02 | 2.37±0.02 | 2.69±0.02 | 2.41±0.02 | 2.74±0.02 | 2.42±0.02 | 2.51±0.02 |

Fig. 2A clearly indicates that hydrolysis of zirconium *n*-propoxide in aqueous media without application of ultrasound results in the formation of zirconia gels with a rather high surface fractal dimension ($D_{S1}$ = 2.3–2.6), and an increase in precipitation pH leads to an increase in the surface fractal dimension ($D_{S1}$). The formation of zirconia surface fractal aggregates can be explained using well-known data on the mechanism of condensation of hydrous silica monomers [62]. It was established that under acidic conditions, condensation occurs mainly between silanol groups located at the ends of polymeric chains, leading to the formation of linear polymers. Oppositely, under basic conditions, the condensation reaction will preferentially occur between the ends and the middle of polymeric chains, thus leading to the formation of more ramified aggregates with a larger fractal dimension. These suppositions are in line with the suggestion that during zirconium alkoxide hydrolysis, condensation can be prevented by the introduction of acid [63].

We can see also that drying of zirconia gels synthesized by forced hydrolysis of $Zr(OPr)_4$ does not crush their fractal structure. The opposite situation was observed by Silva *et al.* [64], when amorphous zirconia has been obtained by hydrolysis of zirconium *n*-propoxide in the presence of acetylacetone. By means of small-angle X-ray scattering (SAXS), the authors have shown that wet gels possessed a fractal structure, which was completely destroyed after conventional drying and transition of wet gels to xerogels. The reason for such a difference in our case is probably connected with the formation of a stronger framework of zirconia particles in the case of forced hydrolysis.

During our previous studies of the mesostructure of hydrated zirconia precipitated under acidic, neutral and basic conditions from aqueous solutions of zirconyl nitrate [39], we have shown that $ZrO_2 \cdot xH_2O$ obtained at pH < 6 possesses no surface fractal properties, while $ZrO_2 \cdot xH_2O$ precipitated at pH 6 and higher exhibits a surface fractal structure. Such a difference accounted for the zirconia point of zero charge value, which was reported to be ~6.0–6.7 [65]. The effect of pH on hydrous zirconia properties is also extensively discussed elsewhere [66–68].

Contrariwise, hydrolysis of zirconium propoxide leads to the formation of surface fractal aggregates, even in acidic media (Fig. 2A), and this is in a good agreement with observations made by Ayral *et al.* [69].

Fig. 2 and Table 2 evidence that ultrasonication leads to a dramatic increase in the surface fractal dimension of zirconia xerogels synthesized both in acidic and basic conditions: $D_{S1}$ rises up to ≈3 according to SANS data. This value is the highest possible for surface fractals and corresponds to the most ramified surface. We did not observe any notable dependence of

SANS-derived $D_{S1}$ values on the pH of initial solutions for sonicated samples – they are almost within the error ranges.

According to literature data, sonication can notably increase zirconia polymeric particle formation and growth [45], due to the facilitation of hydrolysis or polycondensation reactions [70]. The collapse of cavitation bubbles results in the formation of microjets in the liquid, which can substantially change the collision frequency of zirconium hydrated oxide colloid particles, their mean free path and the mobility in the media. All these effects surely lead to the changes in the internal structure of aggregates formed. For example, within the diffusion limited aggregation (DLA) model of fractal clusters formation [71], the highest reachable value of the fractal dimension is equal to 2.50. Aggregates formed within the modified DLA model (upon changes in the free path and the probability of particle coalescence) can possess higher fractal dimensions (up to 3) [72]. These conclusions are in line with previously reported observations on the effects of ultrasound on various sol-gel systems, showing that sonication facilitates the formation of a three-dimensional network with high cross-linking of individual chains [73, 74]. Thus the role of ultrasonication is close to the prolonged digestion of hydrous oxides, which also leads to condensation between particles, and formation of a strengthened porous network able to withstand the capillary forces during drying and calcination to high temperatures [75–78]. But in the case of ultrasonication, these processes proceed with substantially higher rates.

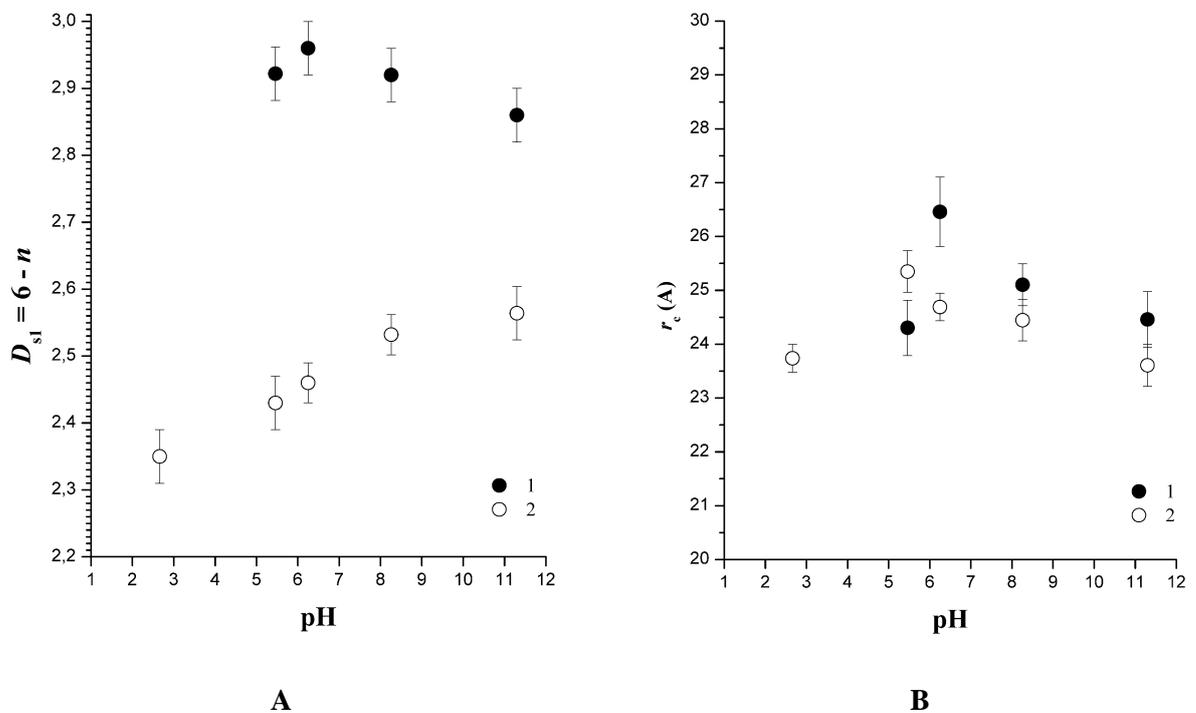

A    B

Fig. 2. Surface fractal dimension $D_{S1}$ (A) and characteristic size of the primary particles $r_c$ (B) derived from SANS data for hydrous zirconia gels precipitated at various pH, both under sonication (1) and without it (2).

Table 2 clearly evidences also that both SANS and SAXS data show a similar effect of sonication on the fractal properties of zirconia ($D_{S1}$), and this corroborates the conclusions made above. The difference between the absolute surface fractal dimension values $D_{S1}$, derived from SANS and SAXS data, is due to the different contrast between X-ray and neutron-scattering density of the solid and that of the pores. Definitely, the X-ray scattering amplitude is directly proportional to the atomic number of an element ($Z$), whereas the neutron-scattering amplitude does not depend on $Z$ [79]. Thus SAXS gives information concerning the distribution of zirconium atoms, while SANS is susceptible to the entire structure and phase boundaries of the xerogel.

Additional evidence that sonication affects the aggregation process is derived from scanning electron microscopy (Fig. 3). One can see that aggregates formed under conventional stirring are nearly spherical with a relatively dense surface (globules), while particles formed in the ultrasonic field are generally shapeless, highly aggregated, and possess a rather developed surface.

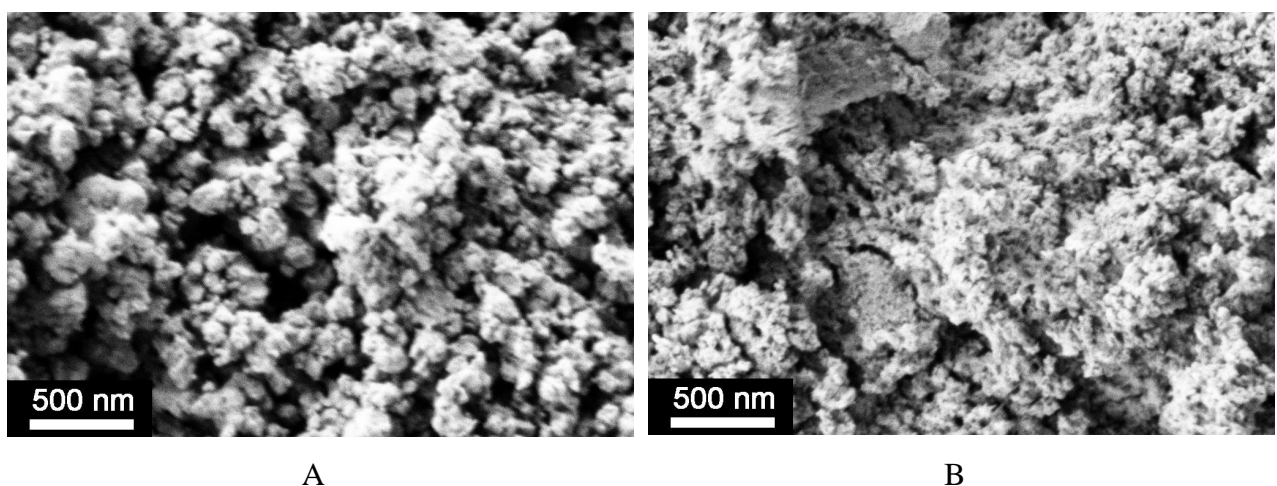

A                                      B

Fig. 3. Scanning electron micrographs of Z-8C (A) and Z-8U (B) samples.

The size of individual hydrated zirconia particles, as estimated from SANS data (Fig. 2B), is nearly independent on precipitation conditions, namely pH of aqueous media and the application of ultrasound. In all the cases, $r_c$ lies within a range of 2.4–2.6 nm. Primary clusters of similar sizes were obtained by Ayral *et al.* [69] for zirconia hydrogels synthesized from acidic media, and by Savii *et al.* [26] for silica gels obtained in US-assisted synthesis. Thus we can conclude that sonication does not affect the size of monomer hydrous zirconia particles, but results in substantial changes in their aggregation patterns.

## 4. Conclusions

In this work, using complementary methods of characterisation of the structure of porous materials on nanometer length scales, namely, low temperature nitrogen adsorption, small-angle neutron scattering and small-angle X-ray scattering, we have demonstrated for the first time how sonication influences the structure of amorphous hydrous zirconia gels formed at various pH. $ZrO_2 \cdot xH_2O$ precipitated from zirconium *n*-propoxide under ultrasonic processing was found to exhibit a more structured surface, characterised by the surface fractal dimension 2.9–3.0, compared to 2.3-2.5 for the non US-assisted synthesis, and possesses a higher specific surface area (about 240 $m^2/g$ compared to 210 $m^2/g$ ), while the sizes of the primary particles remain unchanged. These results are in line with our previous findings [46,47], revealing how ultrasonication affects the mesostucture of amorphous hydrous metal oxide gels.